\documentclass[preprint,12pt]{elsarticle}

\usepackage{graphics}
\usepackage{float}
\usepackage{rotating}
\usepackage{amssymb}
 \usepackage{lineno}

\journal{Astroparticle Physics}

\begin{document}

\begin{frontmatter}



\title{
A statistical procedure for the identification of positrons in the
PAMELA experiment}


\author[a,b]{O. Adriani}
\author[c,d]{G. C. Barbarino}
\author[e]{G. A. Bazilevskaya}
\author[f,g]{R. Bellotti\corref{cor1}}
\ead{roberto.bellotti@ba.infn.it}
\cortext[cor1]{Corresponding
author. Tel: +390805443173}
\author[h]{M. Boezio}
\author[i]{E. A. Bogomolov}
\author[a,b]{L. Bonechi}
\author[b]{M. Bongi}
\author[h]{V. Bonvicini}
\author[j,k,l]{S. Borisov}
\author[b]{S. Bottai}
\author[f,g]{A. Bruno}
\author[g]{F. Cafagna}
\author[d]{D. Campana}
\author[j,d]{R. Carbone}
\author[m]{P. Carlson}
\author[k]{M. Casolino}
\author[n]{G. Castellini}
\author[d]{L. Consiglio}
\author[j,k]{M. P. De Pascale}
\author[k]{C. De Santis}
\author[j,k]{N. De Simone}
\author[j,k]{V. Di Felice}
\author[l]{A. M. Galper}
\author[m]{W. Gillard}
\author[l]{L. Grishantseva}
\author[m]{P. Hofverberg}
\author[h,o]{G. Jerse}
\author[l]{S. V. Koldashov}
\author[i]{S. Y. Krutkov}
\author[e]{A. N. Kvashnin}
\author[l]{A. Leonov}
\author[k]{V. Malvezzi}
\author[k]{L. Marcelli}
\author[p]{W. Menn}
\author[l]{V. V. Mikhailov}
\author[h]{E. Mocchiutti}
\author[f,g]{A. Monaco}
\author[b]{N. Mori}
\author[j,k,i]{N. Nikonov}
\author[d]{G. Osteria}
\author[b]{P. Papini}
\author[m]{M. Pearce}
\author[j,k]{P. Picozza}
\author[q]{M. Ricci}
\author[b]{S. B. Ricciarini}
\author[m]{L. Rossetto}
\author[p]{M. Simon}
\author[j,k]{R. Sparvoli}
\author[a,b]{P. Spillantini}
\author[e]{Y. I. Stozhkov}
\author[h]{A. Vacchi}
\author[b]{E. Vannuccini}
\author[i]{G. Vasilyev}
\author[l]{S. A. Voronov}
\author[m]{J. Wu}
\author[l]{Y. T. Yurkin}
\author[h]{G. Zampa}
\author[h]{N. Zampa}
\author[l]{V. G. Zverev}
\author[r]{D. Marinucci}

\address[a]{
University of Florence, Department of Physics, Via Sansone 1,
I-50019 Sesto Fiorentino, Florence, Italy.}
\address[b]{INFN, Sezione di Florence, Via Sansone 1, I-50019 Sesto Fiorentino, Florence,
Italy.}
\address[c]{University of Naples $"$Federico II$"$, Department of Physics, Via Cintia, I-80126 Naples,
Italy.}
\address[d]{INFN, Sezione di Naples, Via Cintia, I-80126 Naples,
Italy.}
\address[e]{Lebedev Physical Institute, Leninsky Prospekt 53, RU-119991 Moscow,
Russia.}
\address[f]{University of Bari, Department of Physics, Via Amendola 173, I-70126 Bari,
Italy.}
\address[g]{INFN, Sezione di Bari, Via Amendola 173, I-70126 Bari,
Italy.}
\address[h]{INFN, Sezione di Trieste, Padriciano 99, I-34012 Trieste,
Italy.}
\address[i]{Ioffe Physical Technical Institute, Polytekhnicheskaya 26, RU-194021 St. Petersburg,
Russia.}
\address[j]{University of Rome $"$Tor Vergata$"$, Department of Physics, Via della Ricerca Scientifica 1, I-00133 Rome,
Italy.}
\address[k]{INFN, Sezione di Roma $"$Tor Vergata$"$, Via della Ricerca Scientifica 1, I-00133 Rome,
Italy.}
\address[l]{Moscow Engineering and Physics Institute, Kashirskoe Shosse 31, RU-11540 Moscow,
Russia.}
\address[m]{KTH, Department of Physics, and the Oskar Klein Centre for Cosmoparticle Physics, AlbaNova University Centre, 10691 Stockholm,
Sweden.}
\address[n]{IFAC, Via Madonna del Piano 10, I-50019 Sesto Fiorentino, Florence,
Italy.}
\address[o]{University of Trieste, Department of Physics, Via A. Valerio 2, I-34147 Trieste, Italy
.}
\address[p]{University of Siegen, D-57068 Siegen, Germany.}
\address[q]{INFN, Laboratori Nazionali di Frascati, Via Enrico Fermi 40, I-00044 Frascati,
Italy.}
\address[r]{University of Rome $"$Tor Vergata$"$, Department of Mathematics, Via della Ricerca Scientifica 1, I-00133 Rome, Italy.\\
}

\begin{abstract}

The PAMELA satellite experiment has measured the cosmic-ray positron
fraction between 1.5 GeV and 100 GeV. The need to reliably
discriminate between the positron signal and proton background has
required the development of an ad hoc analysis procedure. In this
paper, a method for positron identification is described and its
stability and capability to yield a correct background estimate is
shown. The analysis includes new experimental data, the application
of three different fitting techniques for the background sample and
an estimate of sy\-ste\-ma\-tic uncertainties due to possible
inaccuracies in the background selection. The new experimental
results confirm both solar modulation effects on cosmic-rays with
low rigidities and an anomalous positron abundance above 10 GeV.

\end{abstract}


\end{frontmatter}

\newpage
\section{Introduction}
\label{Introduction} Recent measurements of cosmic-ray electrons and
positrons carried out by the ATIC \cite{atic}, PAMELA \cite{pamela}
FERMI \cite{fermi} and HESS experiments \cite{hess}, extend the
previous balloon-borne
\cite{barwick95,golden_apj,barwick97,boezio2000,aguilar02,beatty04},
and satellite \cite{AMS} measurements and represent a breakthrough
in cosmic-ray physics. In particular it is well known that an
antimatter component that cannot be explained as an effect of a
purely secondary production mechanism, could provide insight into
the nature and distribution of particle sources in our galaxy
\cite{ThThTh}. The PAMELA experiment has reported a measurement of
the positron fraction, i.e. the ratio of positron flux to the sum of
electron and positron fluxes,  $R =
\phi(e^{+})/(\phi(e^{-})+\phi(e^{+}))$, at energies between 1.5 GeV
and 100 GeV, sampled in 16 energy bins. The observations extend the
e\-ne\-rgy range of previous positron measurements and unambiguosly
show an anomalous positron abundance above 10 GeV. A reliable
identification of electrons and positrons has been performed by
combining iformation from independent detectors within the apparatus
\cite{pamela,jint}. The main difficulty in the measurement of R is
the dominating background flux from protons which is $10^3$ (at 1
GV) and $10^4$ (at 100 GV), times the positron flux. Furthermore, a
precise estimate of the proton contamination in the positron sample
is a difficult task.

A widely adopted approach both in
high energy physics and astrophysics, consists of an intensive use
of simulated signal and background samples to train
different multivariate classifiers, such as artificial neural
networks and support vector machines \cite{abazov,ohm}. It
has been demonstrated that such a approach can improve background
rejection in the signal sample \cite{aversa,bbv}. However  this
approach can introduce systematic uncertainties which are difficult to estimate, for the real data.

In this paper we present a method used to obtain an updated the
PAMELA positron fraction \cite{pamela} and further statistical
procedures, based on wavelet and kernel estimates, in order to
estimate the proton contamination in the positron sample. Although
our approach is based on well known statistical techniques, we
believe this methodology can be of interest because the data
analysis is mainly based on the discrimination capabilities of a
single detector, $i.e.$ the electromagnetic calorimeter. Previously
published results \cite{pamela} refer to data collected by the
experiment between July 2006 and February 2008. Here, we present the
methodology applied to larger data set collected between July 2006
and December 2008.

In Section 2  the PAMELA experiment is briefly described. A detailed description of the apparatus
can be found in \cite{pamela_app}. In Section 3 the discriminating variables
used for the analysis are presented. In Section 4 the event
selection procedure is described: this is the first phase of the
analysis and it involves all the detectors of the PAMELA apparatus.
The core of the analysis is described in Section 5. The
methodologies developed to estimate the positron fraction $R$ and
the statistical and systematic uncertainties are illustrated and
applied to the PAMELA data. A summary of the experimental results
and the conclusions are presented in Section 6.

\section{The PAMELA apparatus}
\begin{figure}[H]
%%\centering
\begin{center}
  \includegraphics[height=0.5\textheight]{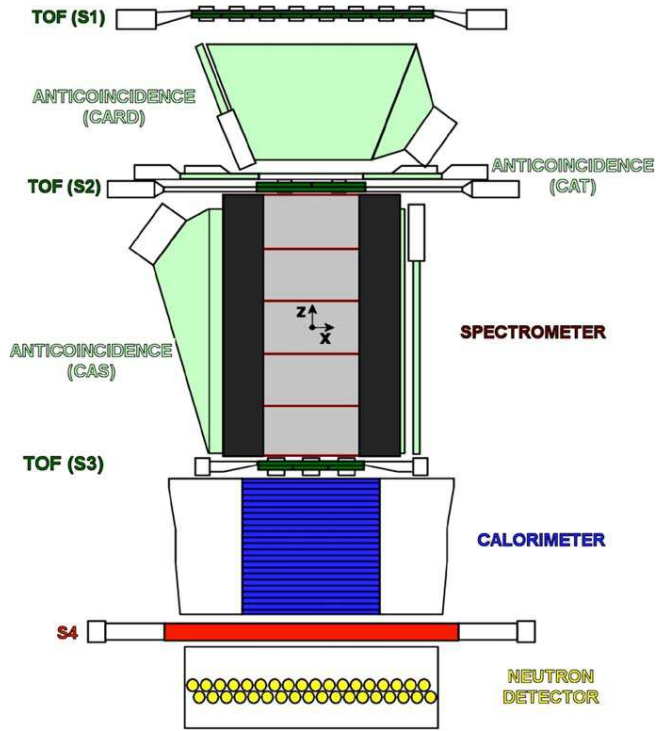}\\
  %oppure prendi flow
  \caption{\footnotesize{A schematic overview of the PAMELA satellite experiment. The
experiment stands $\sim$ 1.3  m high and, from top to bottom,
consists of a time-of-flight (ToF) system (S1, S2, S3 scintillator
planes), an anticoincidence shield system, a permanent magnet
spectrometer (the magnetic field runs in the y-direction), a
silicon-tungsten electromagnetic calorimeter, a shower tail
scintillator (S4) and a neutron detector. The experiment has an
overall mass of 470 kg. }}\label{fig01}
  \end{center}
\end{figure}

As shown in Fig \ref{fig01}, the PAMELA apparatus is composed by the
following detectors (from top to bottom):
\begin{enumerate}
\item a time-of-flight system (ToF (S1, S2, S3));\\
\item a magnetic spectrometer;\\
\item  an anticoincidence system (AC (CARD, CAT, CAS));\\
\item an electromagnetic imaging calorimeter; \\
\item a shower tail catcher scintillator (S4) and\\
\item a neutron detector. \\
\end{enumerate}

The ToF system provides a fast signal for triggering the data
acquisition and measures the time-of-flight and ionization energy
losses (dE/dx) of traversing particles. It also allows down-going
particles to be reliably identified. Multiple tracks, produced in
interactions above the spectrometer, are rejected by requiring that
only one strip of the top ToF scintillator (S1 and S2) layers
register an energy deposition ('hit'). Similarly no hits were
permitted in either top scintillators of the AC system (CARD and
CAT). The magnetic spectrometer consists of a 0.43 T permanent
magnet and a silicon microstrip tracking system. It measures the
rigidity of charged particles through their deflection in the
magnetic field. During flight the spatial resolution is observed to
be 3 $\mu$m in the bending view, corresponding to a Maximum
Detectable Rigidity (MDR), defined as a $100\%$ uncertainty in the
rigidity determination,  exceeding 1 TV.  The dE/dx losses measured
in S1 and the silicon layers of the magnetic spectrometer were used
to select minimum ionizing singly charged particles (mip) by
requiring the measured dE/dx to be less than twice that expected
from a mip. The sampling calorimeter comprises 44 silicon sensor
planes interleaved with 22 plates of tungsten absorber. Each
tungsten layer has a thickness of 0.26 cm corresponding to 0.74
radiation lengths. A high dynamic-range scintillator system (S4) and
a neutron detector are mounted under the calorimeter. The apparatus
is approximately 130 cm tall and with a mass of about 470 kg and it
is inserted inside a pressurized container attached to the Russian
Resurs-DK1 satellite \cite{pamela_app}.

\subsection{The imaging calorimeter}

In this analysis the PAMELA silicon-tungsten sampling imaging
calorimeter \cite{pamela_calo} plays a key role, due to its
capability to give an accurate topological description of the
showers generated by the interaction of the cosmic-ray particles.

Electromagnetic calorimeters have been widely used for particle
discrimination in balloon-borne cosmic-ray experiments
\cite{barwick95,barwick97,golden_apj_94,boezio_apj}. The PAMELA an
imaging calorimeter is evolution of the instrument used in
several balloon-borne experiments
\cite{boezio_apj,bellotti_app_97,golden_apj} and its performances
have been throughly investigated by means of test beam data and Monte
Carlo simulations \cite{pamela_calo}. It is $16.3$ radiation lengths
($0.6$ nuclear interaction lengths) deep, so both electrons and
positrons develop a well-contained electromagnetic shower in the
energy range of interest. In contrast, the majority of the protons
will either pass through the calorimeter as a minimum ionizing
particle or interact deeply in the calorimeter. In fact there is a
high probability ($>$89$\%$) that an electromagnetic shower will
start in the first 3 planes of the calorimeter. For hadronic
showers, the starting point is distributed more uniformly. Particle identification based on the total measured energy and the
starting point of the reconstructed shower in the calorimeter can be
tuned to reject 99.9$\%$ of the protons, while selecting more than
95$ \%$ of the electrons or positrons. The remaining proton
contamination in the positron sample can be eliminated using
additional topological information, including the lateral and
longitudinal profile of the shower. Using particle beam data
collected at CERN it was previously shown that less than one proton
out of $100,000$ passes the calorimeter electron selection up to 200
GeV/c, with a corresponding electron selection efficiency of $80\%$
\cite{pamela_calo}.

\section{Discriminating variable selection}
The misidentification of electrons and protons are the largest
sources of background when estimating the positron fraction. This
can occur if the sign-of-charge is incorrectly assigned from the
spectrometer data, or if electron- and proton-like interaction
patterns are confused in the calorimeter data. The
proton-to-positron flux ratio increases from approximately $10^3$ at
$1$ GeV to approximately $10^4$ at $100$ GeV and represents the
major source of contamination. Robust positron identification is
therefore required and the residual proton background must be
carefully assessed. To do this a single discriminating variable is
considered: the fraction $\mathcal{F}$ of calorimeter energy
deposited inside a cylinder of radius $0.3$ Moli\`{e}re radii. Fig.
\ref{fig1} shows $\mathcal{F}$ as a function of deflection
($\mbox{rigidity}^{-1}$). The axis of the cylinder is defined by
extrapolating the particle track reconstructed in the spectrometer.
The Moli\`{e}re radius is an important quantity in calorimetry as it
quantifies the lateral spread of an electromagnetic shower (about
$90\%$ of the shower energy is contained in a cylinder with a radius
equal to $1$ Moli\`{e}re radius), and depends only on the absorbing
material (tungsten in this case). The events shown in Fig.
\ref{fig1} were selected requiring a match between the momentum
measured by the tracking system and the total detected energy and
the starting point of the shower in the calorimeter. For
negatively-signed deflections, electrons are clearly visible as a
horizontal band with $\mathcal{F}$ lying mostly between $0.4$ and
$0.7$. For positively-signed deflections, the similar horizontal
band is naturally associated to positrons, with the remaining
points, mostly at $\mathcal{F} < 0.4$, designated as proton
contamination. The validity of such event characterization was
confirmed using the neutron yield from the calorimeter and the
ionization ($dE/dx$) losses measured in the spectrometer
\cite{jint}. The spillover limit for positrons is estimated from
particle beam tests to be approximately 300 GeV. From particle beam
tests the spillover limit for positions is estimated to be
approximately 300 GeV, primarily due to the tracker resolution. The
electron spillover background between 1.5 and 100 GeV is negligible.

\section{Event selection}

\begin{figure}[h]
\begin{center}
  \includegraphics[height=0.4\textheight]{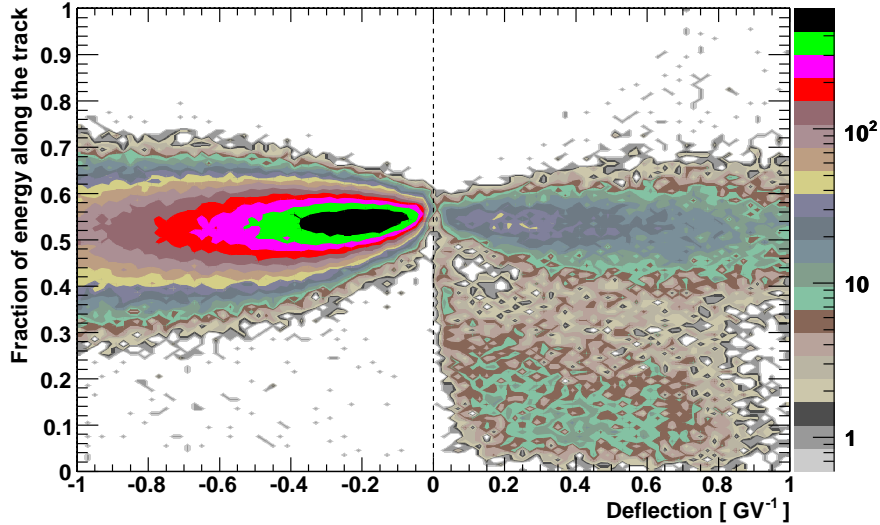}\\
  \caption{\footnotesize{Calorimeter energy fraction $\mathcal{F}$. The fraction of calorimeter energy deposited inside a
cylinder of radius 0.3 Moli\`{e}re radii, as a function of
deflection. The number of events per bin is shown in different
colours, as indicated in the colour scale. The axis of the cylinder
is defined by extrapolating the particle track reconstructed by the
spectrometer. The events were selected requiring a match between the
momentum measured by the tracking system and the total detected
energy and requiring that the electromagnetic shower starts
developing in the first planes of the calorimeter.}}\label{fig1}
  \end{center}
\end{figure}

While the distribution shown in Fig. \ref{fig1} presents a clear
positron signature, the residual proton background distribution must
be quantified. It is worthwhile to note that the background
distribution was obtained using the flight calorimeter data and
there was no dependence on simulations. In order to build a
background model, the total calorimeter depth of 22 detector planes
was divided in two non-mutually exclusive parts: an upper part
comprising planes $1-20$, and a lower part comprising planes $3-22$.
The positron component in positively charged events can be
significantly reduced by selecting particles that do not interact in
the first 2 planes because only $2\%$ of electrons and positrons
with rigidities greater than $1.5$ GV pass this condition. This
requirement selects a nearly pure sample of protons entering the
lower part of the calorimeter (planes $3-22$). The event selection
methodology was further validated using particle beam data collected
prior to lunch and data generated using the PAMELA Collaboration's
official simulation program. This simulation is based on the GEANT
package \cite{gpamela} version 3.21 and reproduces the entire PAMELA
apparatus.

Calorimeter variables (e.g.
total detected energy, and lateral shower spread) were evaluated for
the upper and lower parts of the calorimeter. Electrons and
positrons were identified in the upper part of the calorimeter
using the total detected energy and the starting point of the
shower. As an example Fig. \ref{fig9} shows the energy fraction
$\mathcal{F}$, for negatively charged particles in the rigidity
range $28-42$ GV selected as electrons in the upper half of the
calorimeter (panel a). Panels (b) and (c) show the $\mathcal{F}$
distributions for positively-charged particles obtained for the
lower (upper) part of the calorimeter, i.e. protons (protons and
positrons). The distributions in panels (a) and (b) are clearly
different while panel (c) shows a mixture of the two distributions,
which strongly supports the positron interpretation for the
electron-like $\mathcal{F}$ distribution in the sample of positively
charged events.

\section{Positron/proton discrimination}
\begin{figure}[h]
%\centering
%\begin{center}
  \includegraphics[height=0.45\textheight]{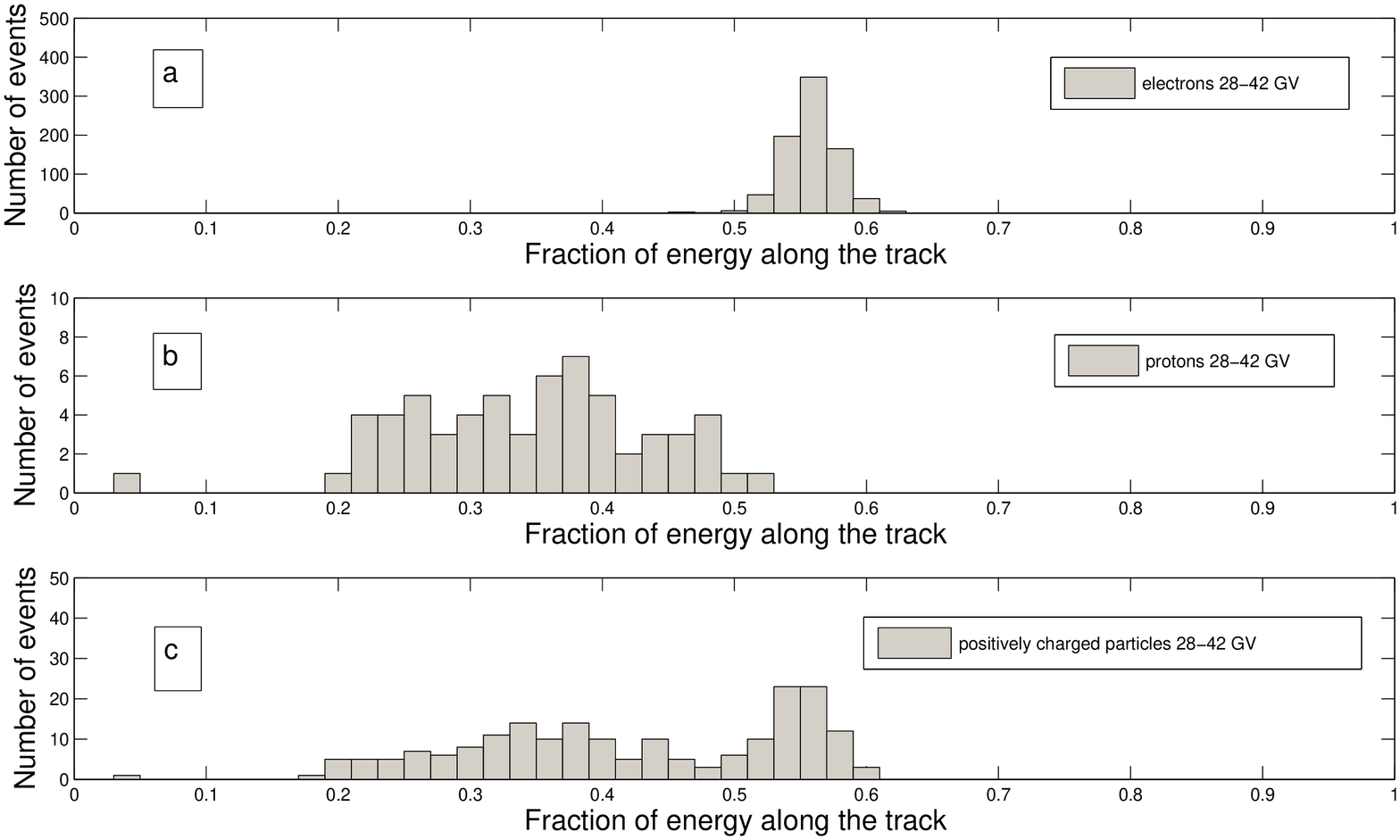}\\
  \caption{\footnotesize{Calorimeter energy fraction $\mathcal{F}$: $28-42$ GV. Panel (a) shows the distribution of the
energy fraction for negatively charged particles, selected as
electrons in the upper part of the calorimeter. Panel (b) shows the
same distribution for positively charged particles selected as
protons in the bottom part of the calorimeter. Panel (c) shows
positively charged particles, selected in the upper part of the
calorimeter, i.e. protons and positrons.}}\label{fig9}
  %\end{center}
\end{figure}

As a result of the event selection described in the last section we
obtained the distributions of pure electrons, pure protons and a
mixture of positrons and protons, as shown in Fig. \ref{fig9}.
Starting from these distributions the determination of the ratio
$R$, with the statistical and systematic error estimates, consists
of four main steps, as summarised in Fig. \ref{fig0}:

\begin{figure}[h]
%%\centering
\begin{center}
  \includegraphics[height=0.5\textheight]{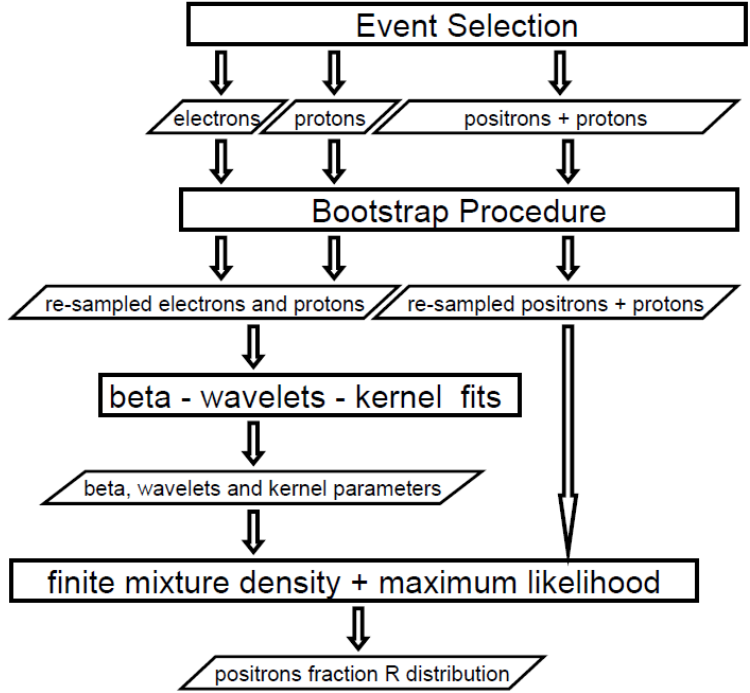}\\
  %oppure prendi flow
  \caption{\footnotesize{Flowchart of the methodology, based on different fitting and bootstrap techniques developed to evaluate the positron fraction $R$.}}\label{fig0}
  \end{center}
\end{figure}

\begin{enumerate}
\item estimation of the probability density functions (pdf) for the experimental distributions shown in Fig. \ref{fig9};\\
\item construction of a finite mixture density of probability;\\
\item estimation of the weight of the mixture by means of the maximum likelihood indicator;\\
\item estimation of the statistical errors by means of a bootstrap procedure.\\
\end{enumerate}
In addition, an estimate of the systematic uncertainties due to
inaccuracies in the background identification is performed.

\subsection{Pdf estimate}

The proton experimental distributions provide information about the
background yields. In order to evaluate these distributions and to
check possible systematic errors in this phase of the analysis,
three different methods have been implemented: beta, wavelets and
kernel.

\subsubsection{Beta pdf}
Since the discriminating variable used for the analysis is the
energy fraction $\mathcal{F}$, spanning the interval [0, 1], the
rational choice is to fit the experimental distribution with a
function spanning in the same interval and with few free parameters,
in order to avoid unphysical modeling of the experimental data. We
used the beta function \cite{beta}:

\begin{equation}\label{beta}
    f(x)=\frac{1}{\beta(p,q)}x^{p-1}(1-x)^{q-1}.
\end{equation}

where $p>0$, $q>0$, and $\beta(p,q)$ is

\begin{equation}\label{int_beta}
    \beta(p,q)=\int_{0}^{1}x^{p-1}(1-x)^{q-1}dx .
\end{equation}

This density has been used to fit both electrons and protons. A set
of parameters for each rigidity bins is obtained and used for the
subsequent steps of the analysis.

\begin{figure}[h]
  \includegraphics[height=0.45\textheight]{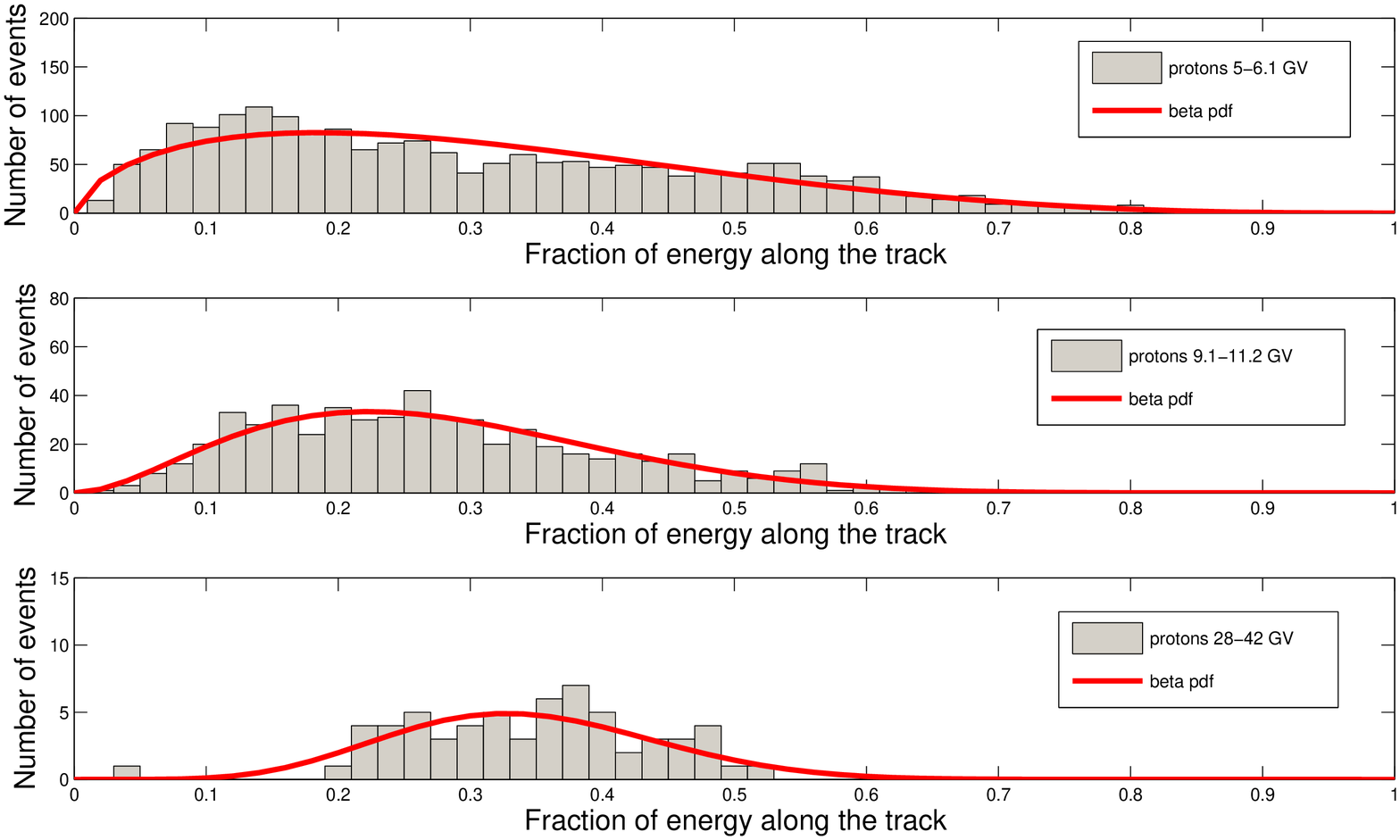}\\
  \caption{\footnotesize{Distribution of the energy fraction for positively charged particles selected as
  protons for 3 different rigidity bins with a fit to a beta pdf. }}\label{fig10}
\end{figure}
The mean of the beta pdf is:
\begin{equation}\label{meanbeta}
   \bar{x}=\frac{p}{p+q}
\end{equation}

and its variance is:

\begin{equation}\label{sdbeta}
    \sigma^2 = \frac{pq}{(p+q)^{2}+(p+q+1)} .
\end{equation}

\subsubsection{Wavelets}

In the previous section, the distribution of the energy fraction was
fitted by means of a fixed family, i.e. we fitted a {\itshape
parametric} law in the statistical jargon. More precisely, we
assumed a priori that the experimental distributions we observed
should be generated according to a specific (beta) law in $[0,1]$.
Although the choice of a beta function is natural for random
variables in this range, it is important to question how much our
final results depend on this assumption, i.e. their degree of
robustness when varying the energy fraction distribution over a much
greater range of possibilities. Our goal here is to explore the
possibility of a {\itshape nonparametric} fit, where there is no a
priori assumption on the energy fraction distribution.
\\
Over the last fifteen years, the statistical literature has focussed
on the estimation of density functions in this broader nonparametric setting. We refer for instance  to \cite{hardle} for an
introduction to this area of research. A wide consensus has formed
on the role of wavelet based methods as the most powerful
statistical techniques for nonparametric density estimation.

\begin{figure}[h]
\centering
\begin{center}
   \includegraphics[height=0.45\textheight]{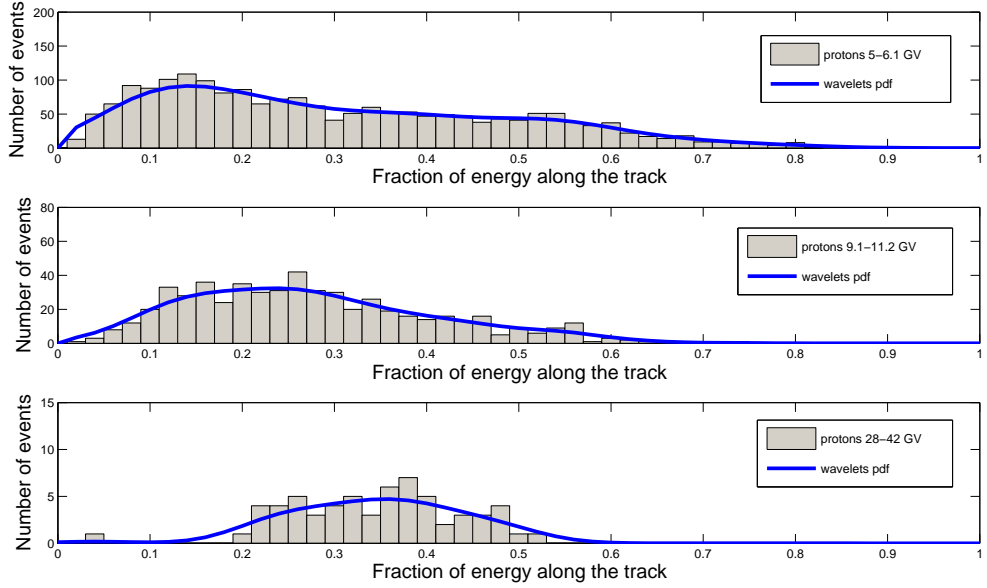}\\
    \caption{\footnotesize{Distribution of the energy fraction for positively charged particles selected as
   protons for 3 different rigidity bins with a wavelets fit.}}\label{fig11}
  \end{center}
\end{figure}

A wavelet system is essentially an orthonormal basis which is
constructed by dilations and translations of a mother and father
function, leading to a multiresolution scheme. The wavelets we are
going to implement are those proposed by Daubechies, which are
computationally convenient as their support in the real domain is
limited. More explicitly, the {\itshape father wavelet} satisfies
\cite{hardle}:

\begin{equation}\label{wave_father}
\varphi(x)=\sqrt{2}\sum_{k}h_k\varphi(2x-k)
\end{equation}
where $h_k$ are suitably chosen weights (\cite{wave_lib}), whereas
the {\itshape mother wavelet} satisfies:

\begin{equation}\label{wave_mat}
\psi(x)=\sqrt{2}\sum_{k}(-1)^{k+1}h_{1-k}\varphi(2x-k)
\end{equation}

The multiresolution expansion of a function $f$ is then provided by:
\begin{equation}\label{wave_f}
f(x)=\sum_{k}\alpha_{k}\varphi_{k}(x)+\sum_{j,k}\beta_{jk}\psi_{jk}(x)
\end{equation}
where $\alpha_{k}$ and $\beta_{jk}$ are approximation and detail
coefficients, respectively, and the elements of the basis are
constructed as

\begin{equation}\label{wave_mat_jk}
\psi_{jk}(x)=2^{j/2}\psi(2^{j}x-k), \ j,k=1,2,...
\end{equation}

 In practice the
coefficients $\alpha_{k}$ and $\beta_{jk}$ are unknown and must be
estimated from the data. Given $x_1,...,x_n$ independent identically
distributed random variables with an unknown density $f$ on
\newcommand{\R}{\mathbb{R}}$\R$, suitable estimators are provided by
\begin{equation}\label{coeffest}
\widehat{\alpha}_{k}=\frac{1}{n}\sum_{i=1}^{n}\varphi_{k}(x_{i}),\
\widehat{\beta}_{jk}=\frac{1}{n}\sum_{i=1}^{n}\psi_{jk}(x_{i}).
\end{equation}

These estimators can be viewed as convolutions of the empirical
histograms of the observations with the elements of the wavelets
basis. At this stage, an obvious estimator may be proposed, by
simply replacing the coefficients $\alpha$, $\beta$ in
(\ref{wave_f}) by their sample estimates. This approach - the so
called {\itshape linear wavelet estimator} - has however been shown
to be suboptimal in general (\cite{hardle} or \cite{wasserman}). On
the other hand, a wide consensus has emerged in the mathematical
statistics community on the use of so-called wavelet thresholding
techniques. Here, {\itshape small} coefficients are suppressed by
introducing a threshold. In particular, in this paper a {\itshape
hard thresholding rule} is used. In this case, the estimator for $f$
is defined by \cite {hardle}:

\begin{equation}\label{wave_thre}
\widehat{f}_n(x)=\sum_{k}\widehat{\alpha}_{k}\varphi_{k}(x)+\sum_{j,k}\widehat{\beta}_{jk}^{H}\psi_{jk}(x)
\end{equation}
where the coefficients $\widehat{\beta}_{jk}^{H}$ are defined by:
\begin{equation}\label{hard_thr1}
\widehat{\beta}_{jk}^{H}=\widehat{\beta}_{jk}I(|\widehat{\beta}_{jk}|>t)
\end{equation}
(the indicator function is defined as usual, e.g. $I(|X|>t)=1$ if
$|X|>t$ , $0$ otherwise). The threshold level is chosen to be
\begin{equation}\label{thr}
t=c\sqrt{\frac{logn}{n}}
\end{equation}
where $c>0$ is a suitably chosen constant, and $n$ is the number of
observations in our sample. Intuitively, the rationale behind these
techniques can be explained as follows. The smaller sample
coefficients can be expected to be largely dominated by noise, so
dropping them will improve the global performance of the estimates.
These argument can be made rigorous, in particular it can be shown
that wavelet thresholding estimators yield basically the optimal
rate of convergence over a wide variety of loss functions, i.e.,
they (nearly) minimize over a wide class of functions $f$ and norms
$\|.\|_{L^p}$ the maximum risk

\begin{equation}\label{minimax}
\R(\hat{f}_n,f)=\max_{f}\langle\|\hat{f}_n-f\|_{L^p}^{p}\rangle
\end{equation}

In practice, this means wavelet thresholding techniques enjoy
robustness properties which are important in our context. They are
sensitive at the same time to large scale features of the unknown
energy distribution, and they are also expected to detect the
possible existence of small scale effects, such as local density
spikes which could affect the final result. We refer again to
\cite{hardle} and \cite{wasserman} for further details and
discussion.

To fit the proton sample, the wavelet thresholding technique with
the Daubechies' basis (in particular db3) has been used. A critical
step has been to find the best value for the parameter $c$ in
(\ref{thr}). We started fixing $c=3$, which is often recommended as
rule-of-the-thumb choice. We further verified by numerical
experiments that our results are very stable for a wide range of
fluctuations around this value. The positron-to-electron ratio
estimates and corresponding confidence intervals are very close
(indeed, in some cases nearly undistinguishable) from those obtained
with the parametric fit of the beta distribution.

\subsubsection{Kernel estimate}
The kernel estimate is a statistical technique used to obtain an
unbinned and nonparametric estimate of the probability density
function. In the univariate case, the general kernel estimate of the
parent distribution is given by \cite{kram}:
\begin{equation}
f(x) = \frac{1}{nh}\sum_{i=1}^{n}K(\frac{x - x_i}{h})
\end{equation}
where $x_{i}$ represents the data and \emph{h} is the smoothing
parameter (also called the bandwidth). It is important to note that
$f(x)$ is bin-independent regardless of choice of $K$. $K$ has the
role to distribute the contribution of each data point in the
evaluation of the probability density function. Istead $h$ have the
task to set the scale of kernel.
\begin{figure}[h]
  \includegraphics[height=0.45\textheight]{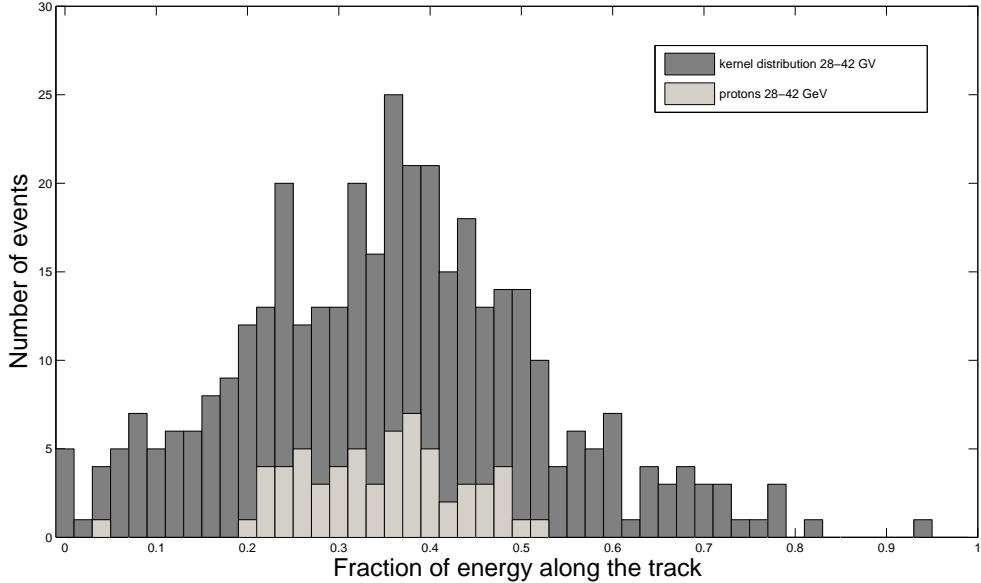}\\
  \caption{\footnotesize{The distribution of positively charged particles selected as
protons (grey) for the rigidity bin $28-42$ GV and the same
distribution modified using the kernel method
(dark-grey).}}\label{fig3}
\end{figure}

Since the discriminating variable is defined in the bounded interval
$[0,1]$ a beta kernel has been used \cite {kernel_beta}. The beta
kernel is a non-negative kernel and it is usually considered to
estimate probability density functions with compact supports.

The number of beta functions generated was equal to the number of
bins of the histogram and each beta function had a mean equal to the
center of histogram bins. The standard deviation of these functions
has been chosen through an application of the Kolmogorov-Smirnov
test so that the initial distribution of protons and the modified
one were statistically compatible, thereby rejecting the null
hypothesis at $5\%$ level.

The beta parameters $(p,q)$ have been calculated inverting
(\ref{meanbeta}) and (\ref{sdbeta}). The kernel bandwith is assumed
to be the histogram bin. The number of events in each histogram of
protons has been increased six fold compared to the original
histogram. Fig. \ref{fig3} shows the real protons and the
pseudo-proton set for rigidity between 28 GV and 42 GV. Each
pseudo-proton sample is then analyzed in the same way as the real
protons (in particular with wavelets-fit), obtaining, for each
energy bin, a new positron fraction.

\subsection{Finite mixture density}
A finite mixture of distributions is used for modelling dataset
extracted from not homogeneous population. It is useful to analize a
sample drawn from an unknown mixture of known distributions. In the
procedure of the finite mixture distributions an experimental
distribution may be approximated as a linear combination of
probability distribution functions (pdfs) \cite {fmd}:
\begin{equation}\label{mist}
g(x,p)=\sum_{i=1}^{n}p_{i}f_{i}(x)
\end{equation}
where $g(x,p)$ is the pdf to estimate, $f_{i}(x)$ are known pdfs,
$n$ is the number of pdfs, $p_i$ are the mixing proportions
($0<p_i<1$ and $\sum_{i=1}^{n}p_{i}=1$) to estimate.

In the present analysis we model, for each energy interval, the
distribution of the calorimeter energy fraction ($\mathcal{F}$) for
positively-charged particles  as mixture distribution \cite{fmd} of
the positrons and protons pdfs:
\begin{equation}\label{pdf_mix}
g(\mathcal{F}) = pf_{b}(\mathcal{F}) + (1-p)f_{s}(\mathcal{F})
\end{equation}
where $f_{b}(\mathcal{F})$ and $f_{s}(\mathcal{F})$ are the
probability density functions for protons and electrons,
respectively and the pdfs $f_{b}$ and $f_{s}$ have been determinated
in the previous section.
As a result of this phase of the analysis a
set of unknown weights $p_j$, with $j$ = 1,...,16, is obtained.

\subsection{Maximum Likelihood}
In order to find the values of unknown weights $p_j$ we used the
well know maximum likelihood method. In the present case the
likelihood function, (for each rigidity bin), is
%replacing (\ref{mist}) in (\ref{like}) is:
\begin{equation}\label{like}
L_j = \prod_{t=1}^{m}\Big[p_jf_{b}(\mathcal{F}_t) +
(1-p_j)f_{s}(\mathcal{F}_t)\Big] .
\end{equation}
where $m$ is the number of independent observations $x_1, x_2,
..., x_m$ in each rigidity bin.

The estimation of the parameters $p_j$ is done by maximizing the
natural logarithm of (\ref{like}):
\begin{equation}\label{max}
   \frac{\partial ln L_j}{\partial p_j} = 0  .
\end{equation}

\begin{figure}[h]
\includegraphics[height=0.45\textheight]{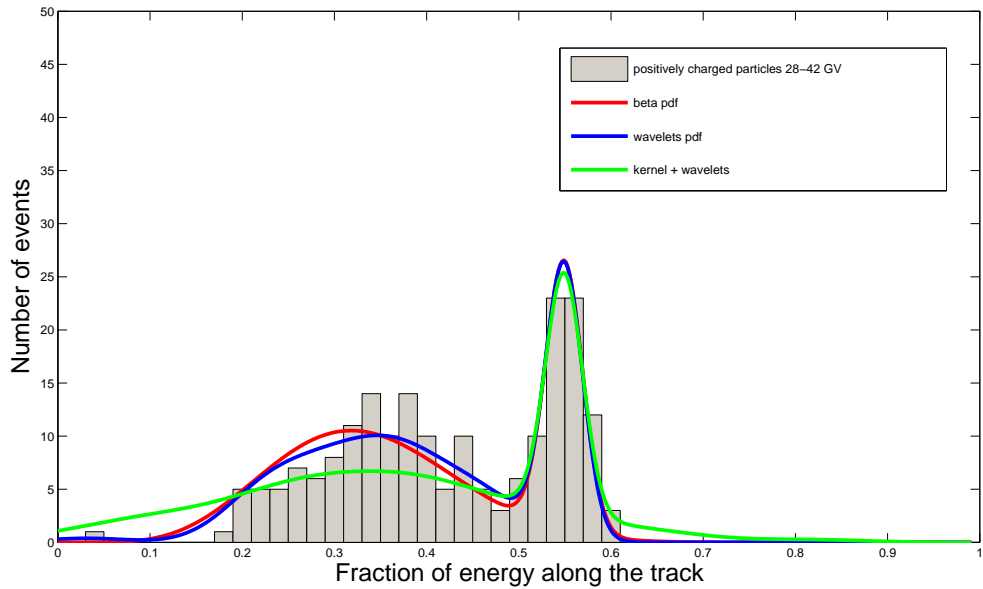}\\
\caption{\footnotesize{The distribution of positively charged
particles for the rigidity bin 28-42 $GV$ showing 3 pdf fits.
}}\label{fig2}
\end{figure}
As a result of the three steps of the analysis, three different
weights of the mixture for each energy bin are obtained. Using the
beta functions, the wavelets transform and the kernel technique. In
the next section the bootstrap technique is introduced. It has been
used to evaluate both the positron fraction $R$ and the statistical
errors of the measurements.

\subsection{Statistical error estimates by means of the bootstrap technique}
The Bootstrap is a powerful method for analyzing small expensive-to-
collect data sets where prior information is sparse \cite{boot1}. In
this method, a set of data is randomly resampled many times with
replacement. Then statistical indicators, such as the standard error
or the confidence interval, are evaluated from these new samples
\cite{boot2}.

This procedure has been used to estimate the statistical error on
the ratio $R$.
\begin{table}[h]

\resizebox*{1\textwidth}{!}{
\begin{tabular}{cccc}
 \hline
Rigidity at  & Percent error  & Percent
error& Percent error \\
spectrometer (GV) &(beta)&(wavelets)&(kernel with wavelets)\\
  %\hline \\
  \hline
%\\
  $1.5-1.8$  & 3.2\% & 2.6\% & 2.6\%\\
  $1.8-2.2$ &2.6\%&2.9\% & 2.6\%\\
  $2.2-2.7$ &2.7\% &2.6\% & 2.6\%\\
  $2.7-3.3$ &2.9\%&3.1\% & 3.1\% \\
  $3.3-4.1$ &3.1\%&3.9\% & 3.9\%\\
  $4.1-5.0$ &3.6\% &3.8\% & 4.3\% \\
  $5.0-6.1$ &3.9\% &5.7\% & 5.3\% \\
  $6.1-7.4$ &4.7\% &4.8\% & 4.4\%  \\
  $7.4-9.1$ &4.9\%&4.9\% & 5.0\% \\
  $9.1-11.2$ & 4.7\% &5.7\% & 5.9\% \\
  $11.2-15.0$ &5.3\% &5.0\% & 5.6\% \\
  $15.0-20.0$ &6.1\% &5.4\% & 6.3\% \\
  $20.0-28.0$ &8.1\% &7.5\% & 8.2\% \\
  $28.0-42.0$ &10.1\% &9.5\% & 11.2\%  \\
  $42.0-65.0$ &13.4\% &12.4\% & 13.0\%  \\
  $65.0-100.0$ &25\% &29.5\% & 25.3\%\\

  \hline
 \end{tabular}
 }
\caption{\footnotesize{Statistical errors on the positron fraction R
for all rigidity bins.}}\label{tab2}
 \end{table}

Each experimental distributions for electrons, protons and
positively-charged particles have been resampled 1000 times, then
the three steps of the analysis procedure previously described have
been repeated. For each rigidity bin, a statistical distribution of
the ratio $R$ is thereby obtained.

As a first step, M = 1000 bootstrap resampling of positives sample
were applied. For each re-sample \emph{i} the unknown parameter
$p_{i}$ was estimated by means of an un binned maximum likelihood
analysis. As a second step the procedure has been repeated $N =1000$
times applying $N$ bootstrap resampling of electron and proton
sample. So $N \times M$ estimations of the number of positron
candidates have been obtained. Then, the final number of positron
candidates was obtained as:
\begin{equation}\label{posit}
\bar{n}=\frac{1}{N}\sum_{i=1}^{N}(\frac{1}{M}\sum_{j=1}^{M}n_{ji})
\end{equation}
where $n_{ji}$ in the number of positron candidates evaluated by
each bootstrap iteration. Therefore also $N \times M$ estimations of
positron fraction have been obtained. In the present analysis we
used the range from the 16th and 84th percentiles of these
distributions as the statistical error estimates of the ratio $R$.
As shown in Table \ref{tab2} the statistical errors on the points
range between $3\%$ and $10\%$ in all bins but the last two and then
increase to just under $30\%$ in the highest energy bin.
\begin{figure}[h]
\begin{center}
\includegraphics[height=0.45\textheight]{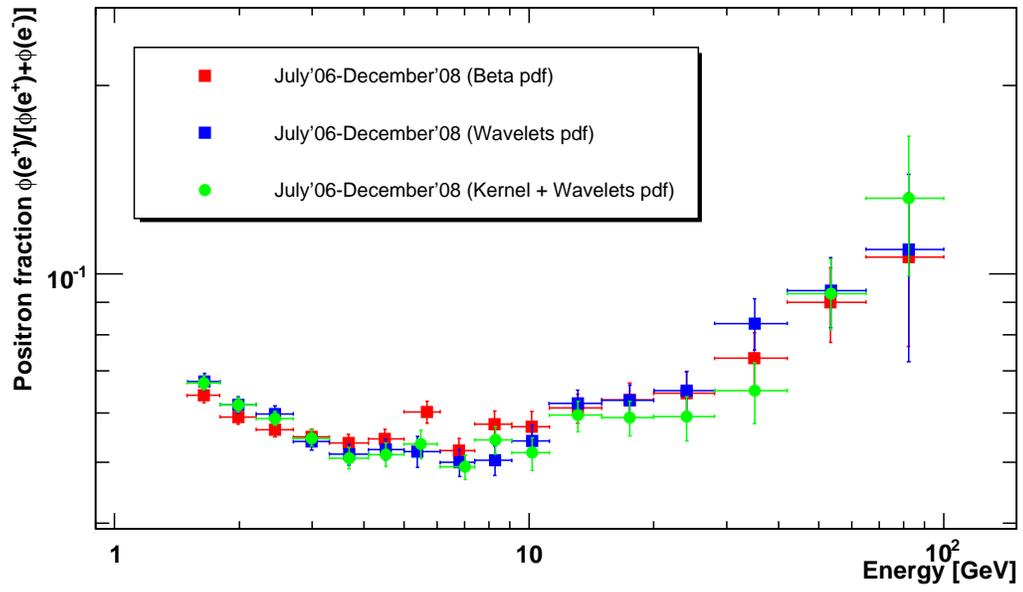}\\
\caption{\footnotesize{The positron fraction R obtained using the
wavelets-fit (blue), beta-fit (red) and kernel with wavelets-fit
(green).}}\label{fig5}
\end{center}
\end{figure}
Fig. \ref{fig5} shows three new estimates of the positron fraction,
using the different fitting techniques adopted in this study.
Moreover, as shown in Tab. \ref{tab4}, the results obtained with the
three different background pdfs are consistent with each other.

\clearpage
\begin{sidewaystable}
\resizebox*{1.2\textwidth}{!}{

\begin{tabular}{ccccc}
  \hline
  Rigidity at &Mean kinetic e\-ner\-gy &Extrapolated $\frac{\phi(e^{+})}{\phi(e^{-})+\phi(e^{+})}$ &Extrapolated $\frac{\phi(e^{+})}{\phi(e^{-})+\phi(e^{+})}$ &Extrapolated $\frac{\phi(e^{+})}{\phi(e^{-})+\phi(e^{+})}$ \\
spectrometer (GV)&at top of payload &at top of payload &at top of
payload
&at top of payload with\\
&beta/wavelets/kernel (GeV)&with beta-fit&with wavelets-fit&kernel with wavelets-fit\\
  %\hline \\
  \hline
\\
  $1.5-1.8$  & 1.65 / 1.65 / 1.65& $0.0639_{-0.0017}^{+0.0017}$ & $0.0673_{-0.0021}^{+0.0021} $ & $0.0670_{-0.0017}^{+0.0017}$ \\
  $1.8-2.2$ &1.99 / 1.99 / 1.99& $0.0591_{-0.0015}^{+0.0015}$ & $0.0618_{-0.0018}^{+0.0018}$ & $0.0618_{-0.0016}^{+0.0016} $ \\
  $2.2-2.7$ & 2.44 / 2.44 / 2.44& $0.0564_{-0.0015}^{+0.0015} $ & $0.0598_{-0.0014}^{+0.0017}$ & $0.0587_{-0.0014}^{+0.0016}$ \\
  $2.7-3.3$ & 2.99 / 2.99 / 2.99& $0.0549_{-0.0016}^{+0.0016} $ & $0.0540_{-0.0017}^{+0.0016} $ & $0.0546_{-0.0015}^{+0.0018} $ \\
  $3.3-4.1$ &3.67 / 3.68 / 3.68& $0.0537_{-0.0017}^{+0.0017} $ & $0.0516_{-0.0021}^{+0.0019}$ & $0.0508_{-0.0020}^{+0.0020} $\\
  $4.1-5.0$ &4.49 / 4.51 / 4.52& $0.0545_{-0.0020}^{+0.0020} $ & $0.0524_{-0.0020}^{+0.0020} $ & $0.0515_{-0.0022}^{+0.0022}$ \\
  $5.0-6.1$ &5.68 / 5.38 / 5.49& $0.0602_{-0.0024}^{+0.0024} $ & $0.0520_{-0.0029}^{+0.0030}$ & $0.0535_{-0.0028}^{+0.0028}$ \\
  $6.1-7.4$ &6.78 / 6.80 / 7.02 & $0.0522_{-0.0024}^{+0.0024}$ & $0.0500_{-0.0025}^{+0.0024}$ & $0.0492_{-0.0022}^{+0.0022} $ \\
  $7.4-9.1$ &8.27 / 8.28 / 8.30& $0.0576_{-0.0028}^{+0.0028} $ & $0.0504_{-0.0027}^{+0.0027} $ & $0.0543_{-0.0027}^{+0.0027} $ \\
  $9.1-11.2$ & 10.16 / 10.17 /10.18& $0.0570_{-0.0033}^{+0.0033}$ & $0.0541_{-0.0031}^{+0.0032}$ & $0.0518_{-0.0032}^{+0.0029}$\\
  $11.2-15.0$ &13.11 / 13.12 / 13.13& $0.0611_{-0.0033}^{+0.0032}$ & $0.0619_{-0.0030}^{+0.0032}$&$0.0595_{-0.0035}^{+0.0031}$ \\
  $15.0-20.0$ &17.50 / 17.51 / 17.51&$0.0630_{-0.0039}^{+0.0039} $ & $0.0628_{-0.0033}^{+0.0036}$ & $0.0590_{-0.0036}^{+0.0039}$ \\
  $20.0-28.0$ &23.99 / 24.00 / 24.01& $0.0645_{-0.0052}^{+0.0052}$ & $0.0651_{-0.0051}^{+0.0048}$ & $0.0592_{-0.0051}^{+0.0045}$\\
  $28.0-42.0$ &34.97 / 35.00 / 34.99& $0.0733_{-0.0074}^{+0.0073}$ & $0.0833_{-0.0077}^{+0.0079}$ & $0.0651_{-0.0074}^{+0.0071}$ \\
  $42.0-65.0$ &53.43 / 53.44 / 53.48& $0.090_{-0.013}^{+0.012}$ & $0.094_{-0.012}^{+0.012} $ & $0.093_{-0.012}^{+0.013}$ \\
  $65.0-100.0$ &82.39 / 82.41 / 82.47& $0.106_{-0.030}^{+0.028}$ & $0.109_{-0.037}^{+0.035}$ & $0.132_{-0.033}^{+0.034}$\\
  \hline
 \end{tabular}
}

\caption{\footnotesize{Summary of the positron fraction results for
the beta-fit, wavelets-fit and kernel with wavelets-fit. The errors
are defined by the range between the 16th and the 84th percentiles
in the $R$ distributions.}}\label{tab4}
\end{sidewaystable}

\subsection{Systematic uncertainties due to inaccuracies in the background selection}
The main sources of systematic uncertainties in the
determination of the positron fraction are investigated in the following. Due to the
equivalence of the results obtained in the previous section with the
three different pdfs, the evaluation of the systematic uncertainties
has been performed using only the beta fit.

This is done by introducing a modification in the background
distribution using the weighted bootstrap technique. This
particular technique consists of positive weights applied to each
observation of the dataset \cite{Barbe}. For each rigidity bin,
starting from a proton sample $n(\mathcal{F})$ with mean $\bar{x}$,
two new samples, $n^+(\mathcal{F})$ and $n^-(\mathcal{F})$, are generated:

\begin{enumerate}
\item $n^+(\mathcal{F})$ with mean $\bar{x}^+>\bar{x}$;\\
\item $n^-(\mathcal{F})$ with mean $\bar{x}^-<\bar{x}$.\\
\end{enumerate}

The bootstrap weights are chosen in order to have both
$n^+(\mathcal{F})$ and $n^-(\mathcal{F})$ statistically incompatible
with $n(\mathcal{F})$, according to the Kolmogorov-Smirnov test.
\begin{figure}[H]
\begin{center}
\includegraphics[height=0.45\textheight]{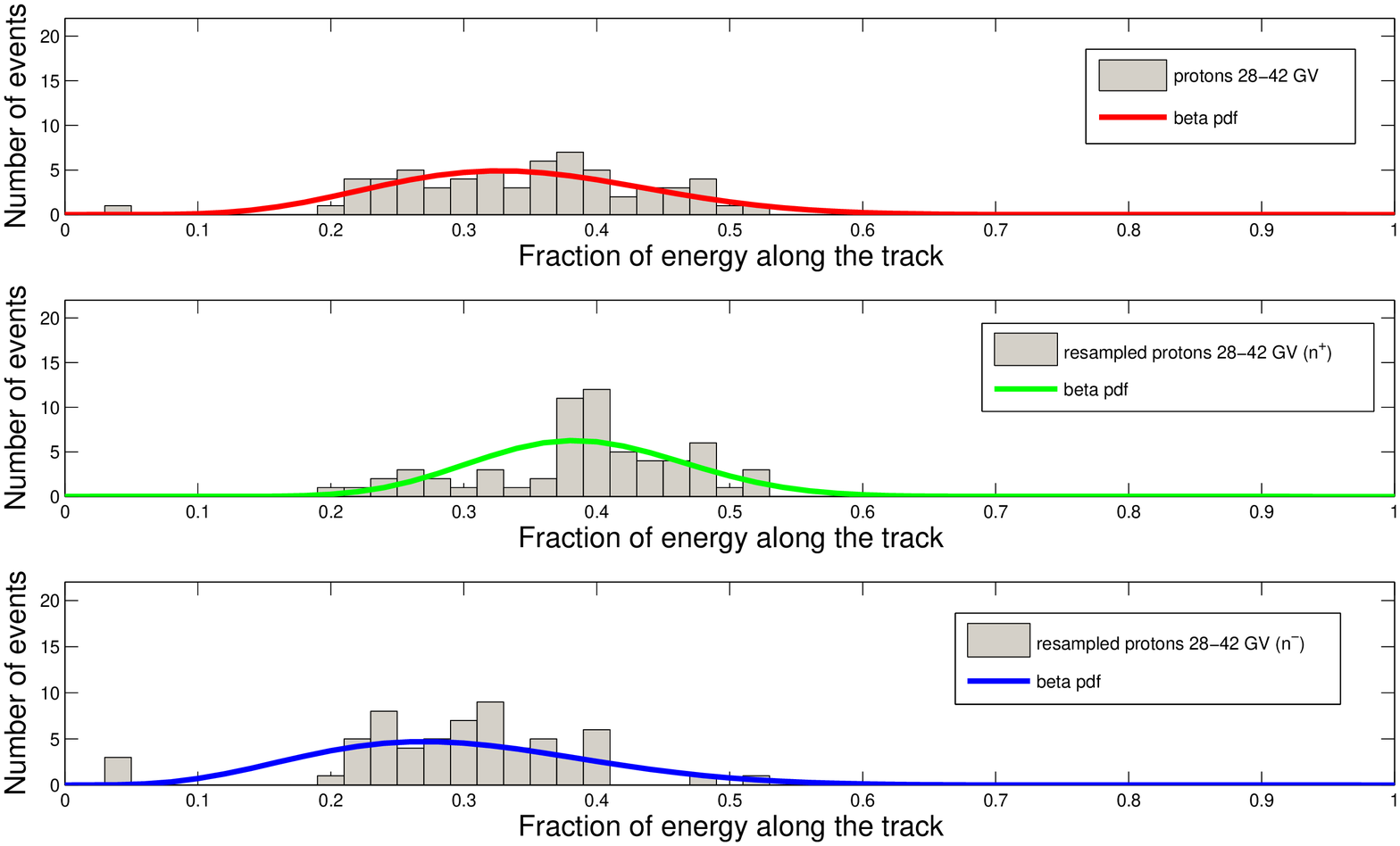}\\
\caption{\footnotesize{The distribution of positively charged
particles selected as protons for the rigidity bin 28 - 42 GV and
the same distribution when modified using the weighted bootstrap
technique.}}\label{fig_sist}
\end{center}
\end{figure}
Fig. \ref{fig_sist} shows protons for the rigidity bin 28 - 42 GV
and the same distribution when modified using the weighted bootstrap
technique. The range encompassing $R-R^+$ and $R+R^-$ is assumed as
an estimate of the systematic uncertainty due the inaccuracies in
the background selection. Tab. \ref{tab5} reports the systematic
uncertainties assesed for each rigidity bin.
\begin{table}[H]
\resizebox*{1\textwidth}{!}{
\begin{tabular}{cccc}

  \hline
 Rigidity at &Mean kinetic e\-ner\-gy &Extrapolated $\frac{\phi(e^{+})}{\phi(e^{-})+\phi(e^{+})}$ &Systematic \\
spectrometer (GV)&at top of payload &at top of payload &uncertainties\\
&beta-fit (GeV)&with beta-fit&\\

  \hline
\\
  $1.5-1.8$  & 1.65 & $0.0639_{-0.0017}^{+0.0017}$ &$_{-0.0017}^{+0.0010 }$  \\
  $1.8-2.2$ &1.99 & $0.0591_{-0.0015}^{+0.0015} $ & $_{-0.0018}^{+0.0011 }$ \\
  $2.2-2.7$ & 2.44& $0.0564_{-0.0015}^{+0.0015}$ &$_{-0.0014}^{+0.0012 }$   \\
  $2.7-3.3$ & 2.99& $0.0549_{-0.0016}^{+0.0016}$ & $_{-0.0013}^{+0.0012 }$  \\
  $3.3-4.1$ &3.67 & $0.0537_{-0.0017}^{+0.0017} $ &$_{-0.0013}^{+0.0011 }$ \\
  $4.1-5.0$ &4.49& $0.0545_{-0.0020}^{+0.0020} $ &$_{-0.0014}^{+0.0018 }$  \\
  $5.0-6.1$ &5.68 & $0.0602_{-0.0024}^{+0.0024} $ & $_{-0.0015}^{+0.0024 }$\\
  $6.1-7.4$ &6.78& $0.0522_{-0.0024}^{+0.0024}$ & $_{-0.0016}^{+0.0024 }$\\
  $7.4-9.1$ &8.27&$0.0576_{-0.0028}^{+0.0028} $ & $_{-0.0018}^{+0.0038 }$ \\
  $9.1-11.2$ & 10.16 & $0.0570_{-0.0033}^{+0.0033} $ &$_{-0.0019}^{+0.0028 }$ \\
  $11.2-15.0$ &13.11& $0.0611_{-0.0033}^{+0.0032}$ &$_{-0.0018}^{+0.0028}$ \\
  $15.0-20.0$ &17.50 &$0.0630_{-0.0039}^{+0.0039} $ & $_{-0.0020}^{+0.0033}$ \\
  $20.0-28.0$ &23.99& $0.0645_{-0.0052}^{+0.0052}$ & $_{-0.0030}^{+0.0045}$ \\
  $28.0-42.0$ &34.97& $0.0733_{-0.0074}^{+0.0073}$ & $_{-0.0044}^{+0.0057}$ \\
  $42.0-65.0$ &53.43& $0.090_{-0.013}^{+0.012}$ & $_{-0.008}^{+0.013}$ \\
  $65.0-100.0$ &82.39& $0.106_{-0.030}^{+0.028}$ & $_{-0.044}^{+0.037}$ \\
  \hline
 \end{tabular}
}
\caption{\footnotesize{Summary of positron fraction results,
obtained with the beta-fit, including statistical and systematic
errors.}}\label{tab5}
 \end{table}
\section{Experimental results and conclusions}

\begin{figure}[h]
\begin{center}
\includegraphics[height=0.45\textheight]{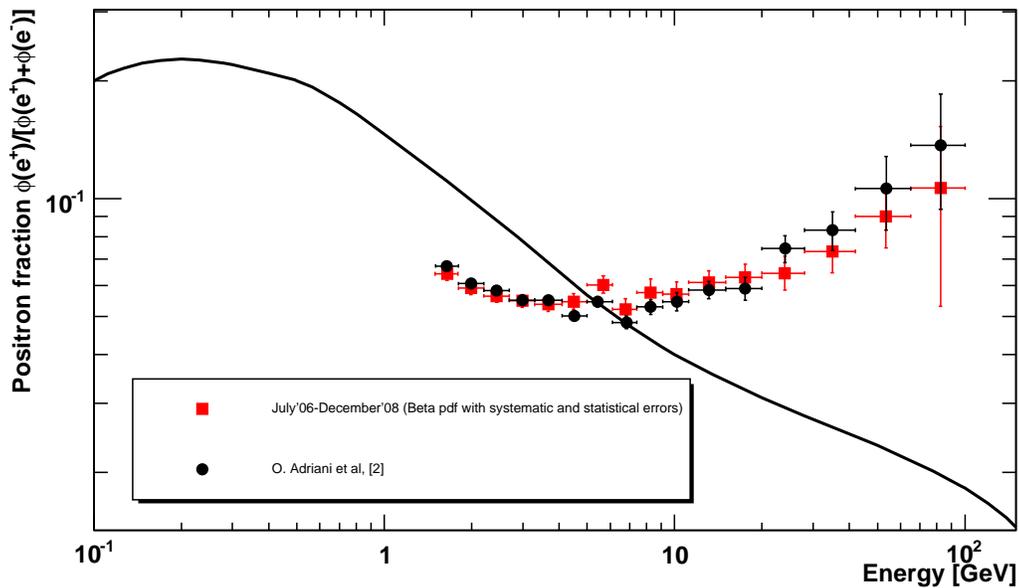}\\
\caption{\footnotesize{The positron fraction R obtained using a
beta-fit with statistical and systematic errors summed in quadrature
(red), compared with the positron fraction reported in \cite{pamela}
(black). The solid line shows a calculation by Moskalenko \& Strong
\cite{moska} for pure secondary production of positrons during the
propagation of cosmic-rays in the galaxy. }}\label{fig_sb}
\end{center}
\end{figure}
Fig \ref{fig_sb} shows the positron fraction R obtained trough
beta-fit with statistical and systematic errors summed in
quadrature, compared with the PAMELA positron fraction previously
reported \cite{pamela}. The solid line shows a calculation by
Moskalenko \& Strong \cite{moska} for pure secondary production of
positrons during the propagation of cosmic-rays in the galaxy.
Proton-positron discrimination is provided the imaging calorimeter,
the capability to yield a trustworthy estimate of the positron and
electron numbers in the cosmic radiation at energies between 1.5 GeV
to 100 GeV has been clearly established. Compared to what is
reported in \cite{pamela}: a) new experimental data, b) the
application of three novel background models and c) an estimate of
the systematic uncertainties has been presented. The new
experimental results are in agreement with what reported in
\cite{pamela} and confirm both solar modulation effects on
cosmic-rays with low rigidities and an anomalous positron abundance
above 10 GeV.

\end{document}